 \def\draftversion{false}
\RequirePackage{ifthen}
\ifthenelse{\equal{\draftversion}{true}}{
  \documentclass[aps,pra,10pt,galley,amsmath,amssymb,nofootinbib,
     superscriptaddress]{revtex4}
}{
  \documentclass[aps,prstab,10pt,twocolumn,amsmath,amssymb,
     superscriptaddress,longbibliography,nofootinbib]{revtex4-1}
}

\usepackage{graphicx}% Include figure files
\usepackage[usenames,dvipsnames]{color} % colors
\usepackage{bm} % bold math
\usepackage{soul} % \st    for strike-out
\usepackage{mathtools}
\usepackage{graphicx}% Include figure files
\usepackage[update,prepend]{epstopdf}
\usepackage{dcolumn} % allow decimal alignment in tables
\usepackage{comment}
\usepackage{gensymb}
\usepackage[linktocpage=true]{hyperref}
\usepackage{hyperref}
\usepackage{comment}
\usepackage{notes2bib}
\hypersetup{
  pdfnewwindow=true, colorlinks=true,
  linkcolor=blue, anchorcolor=blue,
  citecolor=blue, filecolor=blue,
  menucolor=blue, urlcolor=blue}

\ifthenelse{\equal{\draftversion}{true}}{
  \marginparwidth 2.7in
  \marginparsep 0.5in
  \newcounter{comm} % counter for commentaries
  \def\commnext{\stepcounter{comm}}
  \def\commtext{{\bf\color{blue}[\arabic{comm}]}}
  \def\commmar{{\bf\color{blue}[\arabic{comm}]}}
  \def\dvm#1{\commnext\marginpar{\small DV\commmar: #1}\commtext}
  \def\srm#1{\commnext\marginpar{\small SR\commmar: #1}\commtext}
  \def\ssm#1{\commnext\marginpar{\small SS\commmar: #1}\commtext}
  \def\hym#1{\commnext\marginpar{\small HY\commmar: #1}\commtext}
  \def\egm#1{\commnext\marginpar{\small EG\commmar: #1}\commtext}
  \def\mlab#1{\marginpar{\small\bf #1}}
  
  \newcommand{\seclab}[1]{\label{sec:#1}{\Red{\small\;\;[sec:~#1]}}}
  \newcommand{\eqlab}[1]{\Red{\hbox{\small\;\;[#1]}}\label{eq:#1}}
  \newcommand{\figlab}[1]{\Red{\hbox{\small\;\;[fig:~#1]}}\label{fig:#1}}
}{
  \def\dvm#1{}
  \def\srm#1{}
  \def\ssm#1{}
  \def\hym#1{}
  \def\egm#1{}
  \def\mlab#1{}
  
  \newcommand{\eqlab}[1]{\label{eq:#1}}
  \newcommand{\seclab}[1]{\label{sec:#1}}
  \newcommand{\figlab}[1]{\label{fig:#1}}
}

\newcommand{\beq}{\begin{equation}}
\newcommand{\eeq}{\end{equation}}
\newcommand{\bea}{\begin{eqnarray}}
\newcommand{\eea}{\end{eqnarray}}

\newcommand{\eq}[1]{Eq.~(\ref{eq:#1})}

\newcommand{\fref}[1]{Fig.~\ref{fig:#1}}
\newcommand{\Fref}[1]{Figure~\ref{fig:#1}}
\newcommand{\frefs}[2]{Figs.~\ref{fig:#1} and \ref{fig:#2}}

\newcommand{\sref}[1]{Sec.~\ref{sec:#1}}

\newcommand{\Sref}[1]{Section~\ref{sec:#1}}
\newcommand{\ket}[1]{\vert#1\rangle}

\newcommand{\me}[3]{\langle#1\vert#2\vert#3\rangle}

\def\z2{$\mathbb{Z}_2$}

\begin{document}

%===========================%
% TITLE PAGE                %
%===========================%

\title{Electronic structure of Humble defects in Ge and Ge$_{0.8}$Si$_{0.2}$}

\author{Shang Ren}
\affiliation{
Department of Physics and Astronomy, Rutgers University,
Piscataway, New Jersey 08854, USA}

\author{Hongbin Yang}
\affiliation{
Department of Chemistry and Chemical Biology, Rutgers University,
Piscataway, New Jersey 08854, USA}

\author{Sobhit Singh}
\affiliation{
Department of Physics and Astronomy, Rutgers University,
Piscataway, New Jersey 08854, USA}

\author{Philip E. Batson}
\affiliation{
Department of Physics and Astronomy, Rutgers University,
Piscataway, New Jersey 08854, USA}

\author{Eric L. Garfunkel}
\affiliation{
Department of Chemistry and Chemical Biology, Rutgers University,
Piscataway, New Jersey 08854, USA} \affiliation{
Department of Physics and Astronomy, Rutgers University,
Piscataway, New Jersey 08854, USA}

\author{David Vanderbilt}
\affiliation{
Department of Physics and Astronomy, Rutgers University,
Piscataway, New Jersey 08854, USA}

% See definition above
%\mydate

\begin{abstract}

The group-IV diamond-structure elements are known to host
a variety of planar defects, including \{001\} planar defects in C and
\{001\}, \{111\} and \{113\} planar defects in Si and Ge. Among
the \{001\} planar defects, the Humble defect, known for some time
to occur in Ge, has recently also been observed in Si/Ge alloys, but
the details of its electronic structure remain poorly understood.
Here we perform first-principles density functional calculations to
study Humble defects in both Ge and Ge$_{0.8}$Si$_{0.2}$. We also
measure the Si L$_{2,3}$-edge electron energy loss spectra both at
the defect and in a bulk-like region far from the defect, and
compare with theoretical calculations on corresponding Si sites in
our first-principles calculations.  We find that inclusion of
core-hole effects in the theory is essential for reproducing the
observed L$_{2,3}$ edge spectra, and that once they are included, the results
provide a set of fingerprints for different types
of local atomic bonding environments in Ge$_{0.8}$Si$_{0.2}$.  Our
first-principles
calculations reveal that the Humble defects
have a tendency to
enlarge the electronic band gap, which may have potential uses in
band engineering. The use of hybrid functionals for an improved
description of the band gap in these systems is also discussed.

\end{abstract}

\maketitle

%===========================%
% MAIN TEXT                 %
%===========================%

%=================================================
\section{Introduction}
\seclab{intro}
%=================================================

Group IV elements, especially Si and Ge, are now widely used in
semiconductor devices~\cite{paul2004},
optoelectronics~\cite{fadaly2020}, and recently developed quantum
information and computing technologies~\cite{scappucci2020}. Generally,
defects in these materials affect the device properties, with some
even exhibiting useful properties that could be utilized for practical
applications~\cite{queisser1998}. Consequently, the properties of
these defects are of considerable interest for both theoretical and
experimental studies.

One important type of defect is the extended planar defect.  A
well-known example is the \{001\} planar defect in natural
diamond~\cite{lang1964,humble1982,oliver2018}, where recent experimental
work has shown that defect pairs have a zigzag order~\cite{oliver2018}.
In Si, the \{111\}~\cite{jeng1988,muto1991,akatsu2005} and
\{113\}~\cite{takeda1991,dudeck2013} planar defects
are the most common ones.
\{001\} planar defects have also been reported in Si and Ge after hydrogen
implantation~\cite{akatsu2005}. In Ge,
\{001\}~\cite{muto1995,david2007}, \{111\}~\cite{david2007}, and
\{113\}~\cite{ferreiralima1976,david2007,akatsu2005} defects have been reported.
However, only a few of them have been examined using atomic-resolution
high-angle annular dark-field (HAADF) imaging~\cite{dudeck2013,oliver2018},
making it difficult to distinguish between
proposed atomic structures. Alternatively, one
can use electron energy-loss spectroscopy (EELS)
to extract quantitative
information regarding the local atomic bonding
and chemical environments in the vicinity of the
defect, but this has been done only for a few planar defects, e.g.,
for the  \{001\} planar defect in diamond~\cite{oliver2018}.

Several structural models have been proposed to
describe the atomic structure of
\{001\} planar defects.  In 1964, Lang proposed the first model
which assumes that these defects in diamond consist
of nitrogen platelets~\cite{lang1964}. However, the role of
three-fold coordinated nitrogen in the \{001\} planar defects of diamond
is still controversial. 
Later, in 1982, Humble proposed another model in which the
planar defects in diamond consist entirely of
four-coordinated carbon atoms~\cite{humble1982}.
Goss {\it et al.}\ elaborated the
Humble model into five distinct sub-models corresponding to
different arrangements of the atoms residing in the defect layer,
denoted as Humble models (a) to (e)~\cite{goss2001,goss2002,goss2003}.
Even though the Humble model was initially proposed for \{001\} planar
defects in diamond, later work has not confirmed its existence in
diamond, or for that matter, in Si.  Instead,
the Humble defect was observed first in Ge~\cite{muto1995} and
much more recently in a Ge$_{0.8}$Si$_{0.2}$ alloy~\cite{yang2021}.

Planar defects in semiconductors have been the subject of
a variety of computational approaches.
Studies on planar defects in Si have been carried out using molecular
dynamics simulations~\cite{kapur2010,dudeck2013}, empirical potentials and
tight-binding models~\cite{arai1997}, and density-functional theory
(DFT) calculations~\cite{goss2002}. Also, Goss {\it et al.} employed DFT
to study planar defects in diamond~\cite{goss2001, goss2003}.
Although the Humble defects have been experimentally observed in
Ge~\cite{muto1995} and a Ge$_{0.8}$Si$_{0.2}$ alloy~\cite{yang2021},
we are not aware of any
theoretical studies focusing on the electronic properties of
Humble defects in these materials.

In this work, we use first-principles DFT calculations to investigate the
electronic properties of Humble defects in Ge and Ge$_{0.8}$Si$_{0.2}$,
which we denote as GeSi henceforth.
We also carry out experimental measurements of
the Si L$_{2,3}$-edge EEL spectra in the defects and in nearby bulk
regions of GeSi at room temperature, and compare our
theoretical spectra with experiments. Our calculations reveal that
core-hole effects play an important role in describing the Si L$_{2,3}$-edge
EEL spectra of the Humble defects in GeSi.  We also find
that the electronic band gap is locally enhanced in the vicinity of
the Humble defects, potentially offering a unique platform for band
engineering. The use of hybrid functionals to obtain an improved
description of the band gaps in these systems is also discussed.

This paper is organized as follows: In \sref{lang-humble} we briefly
introduce the Lang and Humble models. In \sref{method} we provide
details of our DFT calculations (\sref{method-dft}) and EELS
measurements (\sref{method-eels}). We present the
electronic properties of bulk Ge and GeSe
in \sref{bk-ge-gesi}, and show how the DFT band-gap problem, which is
particularly severe for bulk Ge, can be fixed by
performing hybrid-functional DFT calculations. \Sref{humble-ge-gesi}
provides futher discussion of
the electronic properties of the Humble defects
in Ge and GeSi, focusing on the effect on the local band gap.
The EELS measurements and their comparison with simulations are presented in
\sref{eels}, both in the bulk-like region (\sref{eels-bulk}) and the in
the defect core (\sref{eels-defect}). We summarize our findings and
conclude in \sref{disscon}.

%=================================================
\section{The Lang and Humble models}
\seclab{lang-humble}
%=================================================

\begin{figure}
\centering\includegraphics[width=\columnwidth]{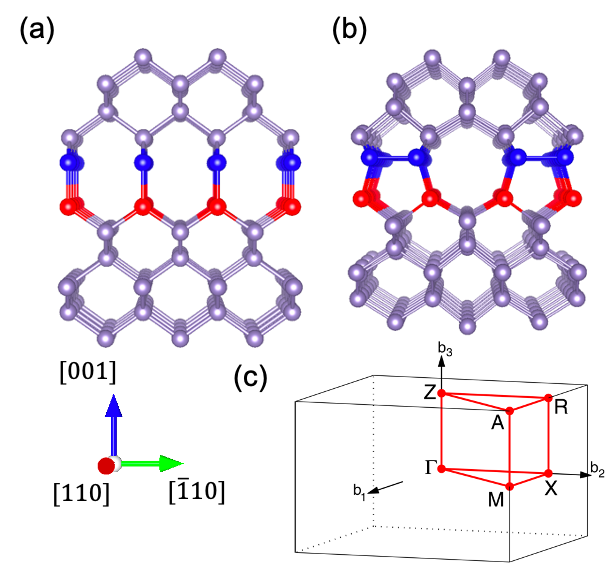}
\caption{Atomic structures of the (a) Lang and (b) Humble models.
Atoms in the top and bottom defect-core layers are colored blue and red.
(c) Brillouin zone for the supercell structures employed in the DFT
calculations for both models; high-symmetry points are labeled.}
\figlab{lang-humble-bz}
\end{figure}

Before presenting the Humble model~\cite{humble1982}, it is
instructive to revisit the Lang model~\cite{lang1964},
which can be regarded as the predecessor of the Humble model.
Lang initially proposed this model to describe the \{001\} planar
defect in diamond.  As shown in \fref{lang-humble-bz}(a), there
are two layers of nitrogen atoms in the Lang model, shown as
red and blue.  The perfect diamond structure can be
recovered by removing one of these two defect layers
and rebonding the atoms with dangling bonds.  We will
use the terminology of ``defect core" to denote the two
layers of defect atoms in the Lang model, and in the Humble model as well.
The atoms in the bulk are all four-coordinated, whereas the atoms in the
defect core are three-coordinated.  If the defect core consists
of column-IV atoms as in the bulk, these atoms would have costly
dangling bonds, an observation that motivated Lang to suggest
trivalent N atoms for the core sites instead.

Unlike the Lang model, atoms in the
Humble defect core are four-coordinated, just as they are in
the bulk.  The Humble model can be derived from the Lang model
by pairing neighboring core atoms and moving them closer
to one another to form a dimer bond, thereby removing two dangling
bonds and converting all atoms to fourfold coordination.
As shown in \fref{lang-humble-bz}(b), the red atoms in the Humble
model are displaced in the $[110]$ direction to form dimers,
while the blue atoms are displaced in the $[\bar{1}10]$ direction,
relative to the Lang model.

Humble's initial proposal assumed one particular pairing
arrangement in the defect core, but other
atomic arrangements are possible. In
Refs.~\cite{goss2001,goss2002,goss2003}, the original Humble model
was extended by proposing five different possible types
of atomic arrangements in the core, denotes as types (a) to (e),
as shown in \fref{humble-a-to-e}, with the original model corresponding
to model (a).  These five Humble models have been studied theoretically
in diamond~\cite{goss2001,goss2003} and Si~\cite{goss2002}.

\begin{figure}
\centering\includegraphics[width=\columnwidth]{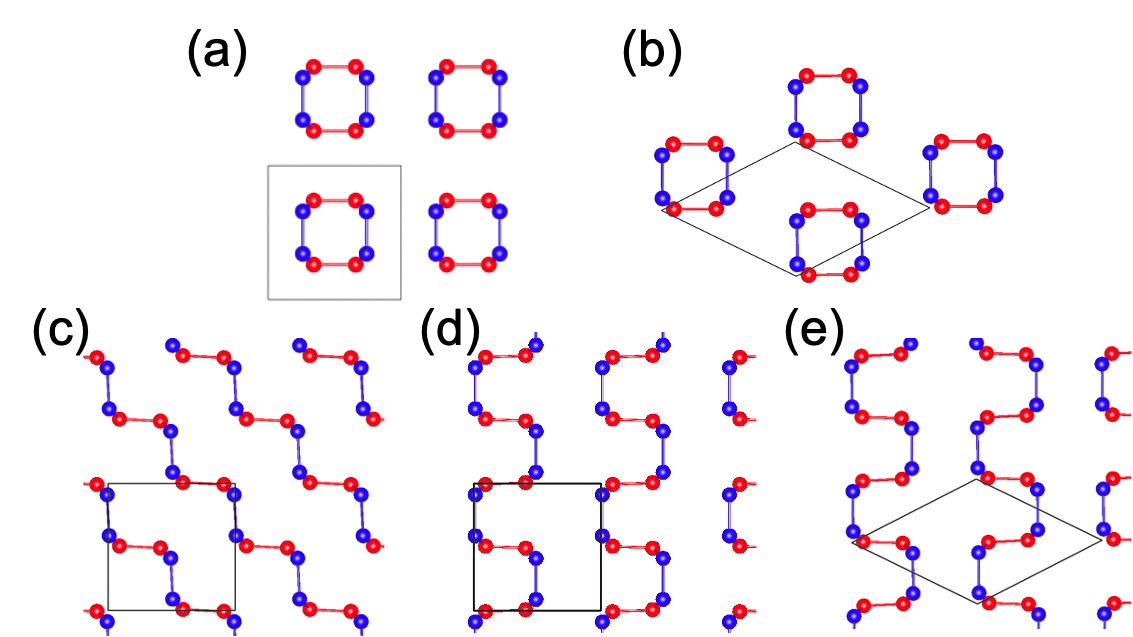}
\caption{Atomic arrangements of Humble defects (a)
to (e), shown in top view using the same color coding as in
\fref{lang-humble-bz}.  Coordinates are those of Ge defects relaxed
using DFT.}
\figlab{humble-a-to-e}
\end{figure}

%=================================================
\section{Methods}
\seclab{method}
%=================================================

%=================================================
\subsection{DFT calculations}
\seclab{method-dft}
%=================================================

All the reported DFT calculations are performed using the
Vienna Ab-initio Simulation Package (VASP)~\cite{kresse1996}
and the projector-augmented wave~\cite{blochl1994,kresse1999}
method with Ge $4s^2 4p^2$ and Si $3s^2 3p^2$ pseudopotential
valence configurations.  Standard DFT calculations employ the
generalized-gradient-approximation (GGA) exchang-correlation
functional of Perdew, Burke, and Ernzerhof (PBE)~\cite{perdew1996},
henceforth denoted as GGA-PBE.
The convergence criteria for forces and energies during
structural relaxation are $10^{-3}\,
\mbox{eV}/\mbox{\AA}$ and $10^{-7}$~eV respectively.
In some cases, band structures and gaps are also computed using the
HSE03 hybrid functional~\cite{heyd-jcp03,heyd-jcp04,heyd-jcp06}. In
those cases, the experimental lattice constants are used in the
calculation.
The cutoff energies for the plane-wave basis set are 500 
and 400\,eV for GGA-PBE and HSE03, respectively.
Other numerical details such as the size of the k-points mesh or
the inclusion of spin-orbit coupling
or whether the hybrid functional is used will be specified below.

The virtual crystal approximation (VCA) as implemented in
Ref.~\cite{bellaiche2000} is used when
simulating the Ge$_{0.8}$Si$_{0.2}$ alloy.
In this approach, every atom is identical, with an identity that is
a mix of 80\% Ge and 20\% Si in the Ge$_{0.8}$Si$_{0.2}$ material.
The VCA takes care of averaging over the ensemble of all possible
distributions in a mean-field sense. 
The AFLOW~\cite{setyawan2010}
online tools are used to analyze the structure, and the {\sc
PyProcar}~\cite{herath2020} package is used for the post-processing
of the electronic structure data.

%=================================================
\subsection{Electron energy loss spectroscopy}
\seclab{method-eels}
%=================================================

The Si L$_{2,3}$ edge EEL spectra are acquired
using a scanning transmission electron microscope equipped with
a electron monochromator. In our experiments, the size of our
electron beam is about 2~$\mbox{\AA}$, small enough to allow
separate imaging of the bulk
and defect regions of Ge$_{0.8}$Si$_{0.2}$. To avoid radiation
damage, a rectangular scan window is placed at the defect or bulk
region during EEL spectra acquisition. We use the monochromator
to improve the energy resolution to about 100 meV and use an
EELS detector dispersion of 25.7\,meV/pixel. After focusing the
zero-loss peak, the spectrometer is further tuned to obtain
optimal focus in the vicinity of 100\,eV, close to the Si L$_{2,3}$
edges of interest. With a detector dwell time of 2\,sec, 30 to
50 EEL spectra are taken in serial and summed to obtain good
statistics. We then fit and subtract an exponential background
from the spectra and
perform a deconvolution to obtain the Si L$_{3}$ edge spectra
presented in \sref{eels}.

%=================================================
\section{Bulk properties of G{\lowercase{e}} and G{\lowercase{e}}$_{0.8}$S{\lowercase{i}}$_{0.8}$}
\seclab{bk-ge-gesi}
%=================================================

\begin{table}
\caption{\label{tab:ge-gesi-bulk} Lattice constants calculated
using GGA-PBE, and band gaps computed from HSE03, for
Ge and Ge$_{0.8}$Si$_{0.2}$.
Experimental lattice constants and band gaps at room temperature
are from Refs.~\cite{dismukes1964} and \cite{braunstein1958}, respectively.
}
\begin{ruledtabular}
\begin{tabular}{ccccc}
& \multicolumn{2}{c}{Lattice constant ($\mbox{\AA}$)} & \multicolumn{2}{c}{Band gap (eV)} \\
& GGA-PBE & Exper. & HSE03 & Exper. \\
\colrule
Ge & $5.78$ & $5.66$ & $0.65$ & $0.66$ \\
Ge$_{0.8}$Si$_{0.2}$ & $5.77$ & $5.61$ & $0.81$ & $0.85$ \\
\end{tabular}
\end{ruledtabular}
\end{table}

We start from bulk Ge and Ge$_{0.8}$Si$_{0.2}$. Due to the
quasi-random distribution of atoms in the Ge$_{0.8}$Si$_{0.2}$
alloy, it is computationally challenging to simulate the
Ge$_{0.8}$Si$_{0.2}$ alloy within DFT.  Here we employ the virtual
crystal approximation (VCA) as implemented in Ref.~\cite{bellaiche2000}
to construct
a virtual atom which is a mixture of 80\% Ge and 20\% Si, and
then build the Ge$_{0.8}$Si$_{0.2}$ structures consisting of the
virtual atoms.  Due to the similarities between the Ge and Si
atoms, we expect the results obtained using the VCA to provide
a reasonable description of the studied system.

First, we use GGA-PBE to relax the bulk Ge and
Ge$_{0.8}$Si$_{0.2}$ diamond structures.  The relaxation
is performed on a two-atom primitive unit cell with a $16
\times 16 \times 16$ Monkhorst-Pack (MP)~\cite{monkhorst1976}
k-mesh. The GGA-PBE optimized 
lattice constants of bulk Ge and Ge$_{0.8}$Si$_{0.2}$ are summarized
and compared with experiment in Table~\ref{tab:ge-gesi-bulk}.
The GGA-PBE calculation slightly overestimates
the lattice constants, but it
severely underestimates the electronic band gap.  The GGA-PBE
predicts both the bulk Ge and Ge$_{0.8}$Si$_{0.2}$ systems
to be semimetals, as shown in \fref{ge-gesi-bulk-band}, although
both these systems are experimentally known to be semiconductors.
We therefore tested all of the local-density approximation (LDA)
and GGA exchange-correlation functionals implemented in VASP, but
confirmed that all of them incorrectly predict
bulk Ge to be a semimetal, in agreement with previous
DFT studies~\cite{igumbor2017hybrid}.  However, hybrid
functionals~\cite{heyd-jcp03,heyd-jcp04,heyd-jcp06} are known to correctly
predict a non-zero band gap in bulk Ge~\cite{deakPRB2010}.  We therefore
adopt the HSE03 hybrid functional~\cite{heyd-jcp03,heyd-jcp04,heyd-jcp06}
for an improved description of band-structure properties in this work.

The HSE03 results for the band gaps are also presented in
Table~\ref{tab:ge-gesi-bulk}.  These are computed at the
experimental lattice constants given in the Table.
Due to the computational expense
of the HSE03 calculations, we use an $8\times 8\times 8$
MP k-mesh for the self-consistent part of the HSE03
calculation with a plane-wave energy cutoff of 400\,eV.
The HSE03 band structures are shown in \fref{ge-gesi-bulk-band}.
An indirect band gap is observed for both Ge and
Ge$_{0.8}$Si$_{0.2}$ and the gap values are in good agreement
with experiment.

\begin{figure}
\centering\includegraphics[width=\columnwidth]{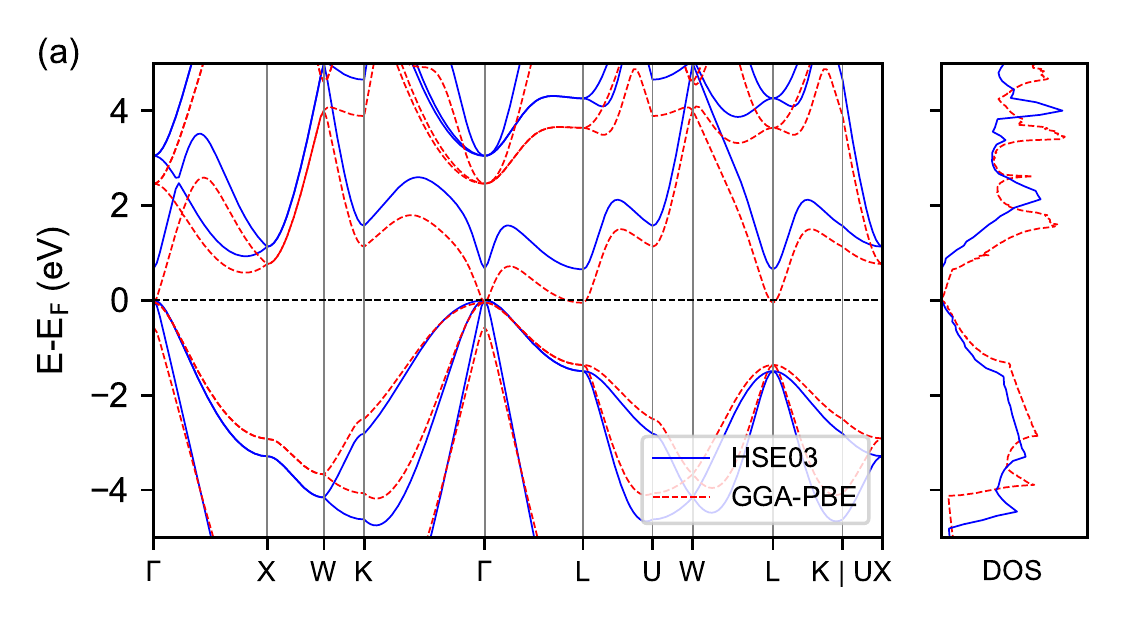}
\centering\includegraphics[width=\columnwidth]{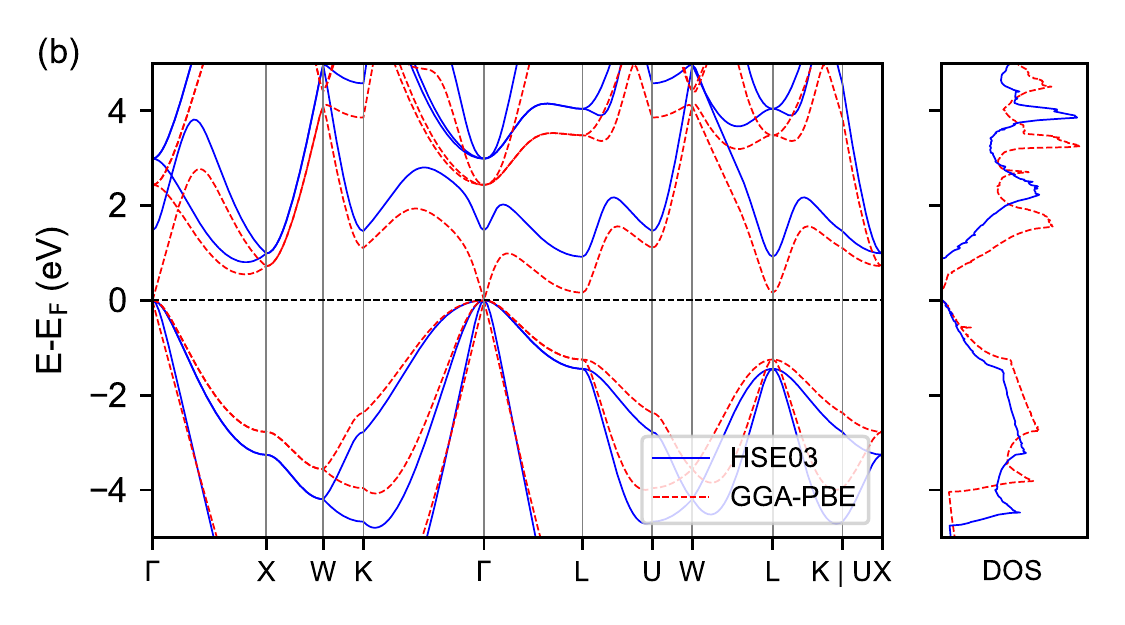}
\caption{The band structure and density of states (DOS) of (a) Ge bulk and (b)
Ge$_{0.8}$Si$_{0.2}$ bulk. In each figure, the left panel is the
band structure and the right panel is the DOS. The red dashed
and blue solid lines
are calculated using the GGA-PBE and HSE03 functionals, respectively.
The Fermi energy coincides with the top of the valence band.}
\figlab{ge-gesi-bulk-band}
\end{figure}

We also calculate the density of states (DOS) using both the
GGA-PBE and HSE03 functionals, as shown in the right panels
of \fref{ge-gesi-bulk-band}.  It is evident that
the GGA-PBE DOS in the conduction-band region almost matches
that of the HSE03 one except for a rigid shift of states along
the energy axis.  This is reasonable because the momentum-space
dispersion of the conduction bands does not change substantially
between the GGA-PBE and HSE03 calculations, as can be seen in
\fref{ge-gesi-bulk-band}. The HSE03 correction to the
bulk Ge and Ge$_{0.8}$Si$_{0.2}$ band structures can thus be
said to be of the ``scissors'' type.

We now discuss the role of Si alloying on the electronic
band structure of bulk Ge.  As shown in \fref{ge-gesi-bulk-band}
and Table.~\ref{tab:ge-gesi-bulk}, the HSE03 band gap is enlarged in
the Ge$_{0.8}$Si$_{0.2}$ alloy compared to pristine Ge.  Both Ge
and Ge$_{0.8}$Si$_{0.2}$ exhibit an indirect band gap.  For Ge
the indirect band gap is from $\Gamma$ to $L$, whereas
for Ge$_{0.8}$Si$_{0.2}$ it is from $\Gamma$ to the
valley near $X$ along $\Gamma$--$X$.  This agrees well with experimental
data obtained using EELS~\cite{batson1991,batson1995} and
another theoretical work performed using nonlocal empirical
pseudopotentials~\cite{fischetti1996}.  Besides the above-mentioned
minor differences, the overall electronic properties of bulk Ge
and Ge$_{0.8}$Si$_{0.2}$ are very similar.

%=================================================
\section{Humble defects in G{\lowercase{e}} and G{\lowercase{e}}$_{0.8}$S{\lowercase{i}}$_{0.2}$}
\seclab{humble-ge-gesi}
%=================================================

The Humble defects are experimentally observed in both
Ge~\cite{muto1995} and Ge$_{0.8}$Si$_{0.2}$. The DFT calculations
show that the Humble (a) defect is
energetically the most favorable one among the five distinct Humble
defects models shown in \fref{humble-a-to-e} ~\cite{yang2021};
for both Ge and GeSi, Humble defects (b-e) are at least 45\,meV
(per interstitial atom) higher in energy,
as reported in Ref.~\cite{yang2021}.
Therefore, here we focus only on the Humble defect (a) in Ge and
Ge$_{0.8}$Si$_{0.2}$ and calculate its electronic properties.
If not specified, the word ``Humble defect" henceforth denotes
the Humble (a) defect.

The Humble structure is built by the procedure mentioned in
\sref{lang-humble}. The supercell consists of 52 atoms in 13
layers, and we assume periodic boundary conditions in all
three dimensions. The space group of the Humble structure is $P\bar{4}m2$ (no.~115).
The relaxation is performed using GGA-PBE
with a $6\times6\times3$ MP grid of k-points. The in-plane
lattice constants ($a$ and $b$) are fixed to the bulk value
obtained from the GGA-PBE relaxation of the defect-free bulk structure,
whereas the supercell lattice constant $c$ is allowed to relax.
The relaxed $c$ lattice constants are reported
for the Ge and Ge$_{0.8}$Si$_{0.2}$ Humble structures in the middle
column of Table.~\ref{tab:ge-gesi-humble}.

\begin{table}
\caption{\label{tab:ge-gesi-humble}
The lattice constants and band gaps of Ge and Ge$_{0.8}$Si$_{0.2}$ Humble structures as computed within the GGA-PBE approximation.}
\begin{ruledtabular}
\begin{tabular}{cccc}
& \multicolumn{2}{c}{Lattice constants ($\mbox{\AA}$)} & Band gap (eV) \\
& $a=b$ & $c$ &    \\
\colrule
Ge & $8.18$ & $19.43$ & $0.13$ \\
Ge$_{0.8}$Si$_{0.2}$ & $8.16$ & $19.39$ & $0.24$  \\
\end{tabular}
\end{ruledtabular}
\end{table}

The calculated electronic band structures of the GGA-PBE
optimized 13-layer Humble structures are presented
in \fref{ge-gesi-humble-band}.  The band structures of the two
Humble structures are very similar, suggesting that the
Ge Humble structure can be used as a good approximation for the
Ge$_{0.8}$Si$_{0.2}$ one.  Surprisingly, our GGA-PBE calculations predict
both the Ge and Ge$_{0.8}$Si$_{0.2}$ Humble structures to be
small-gap insulators, even though the same GGA-PBE calculations
predict defect-free bulk Ge and Ge$_{0.8}$Si$_{0.2}$ to be
semimetallic.  That is, we find that the introduction of Humble
defects in bulk Ge or Ge$_{0.8}$Si$_{0.2}$ tends to open the band
gap.  The computed gaps of the two Humble structures are listed in
Table.~\ref{tab:ge-gesi-humble}.  Notably, the band gap for each
structure is indirect; the valence-band maximum is between the
$A$ to $Z$ point, whereas the conduction-band minimum is at the
$\Gamma$ point of the Brillouin zone.

\begin{figure}
\centering\includegraphics[width=\columnwidth]{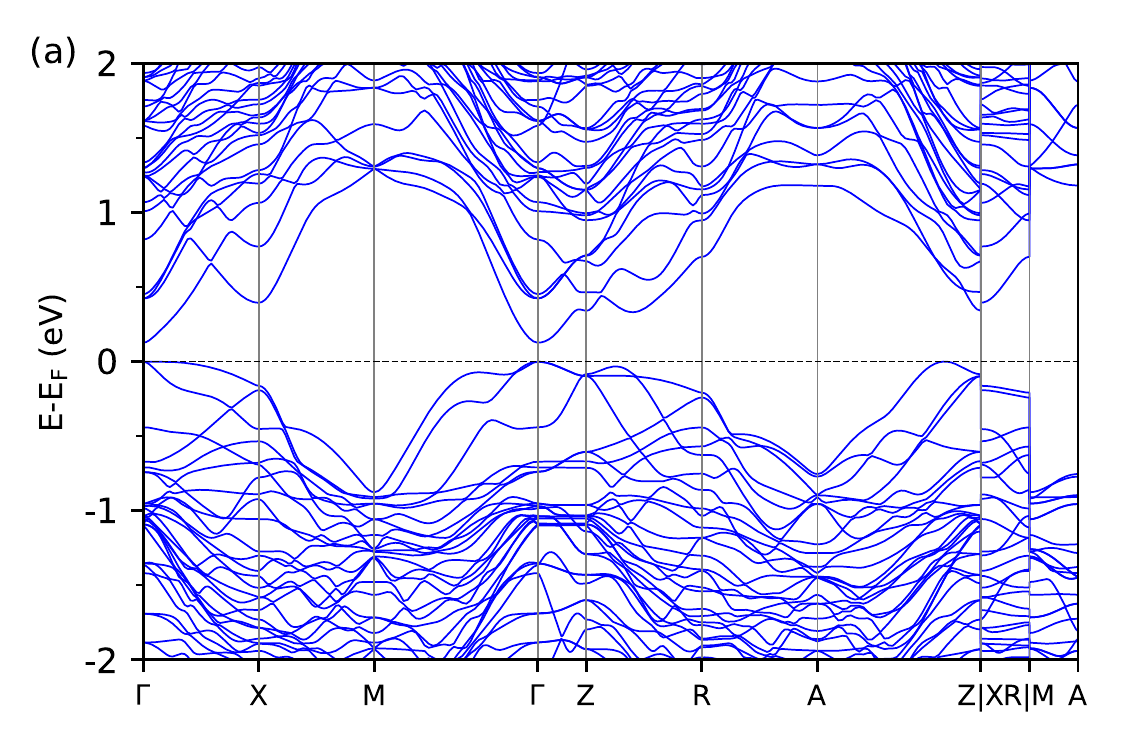}
\centering\includegraphics[width=\columnwidth]{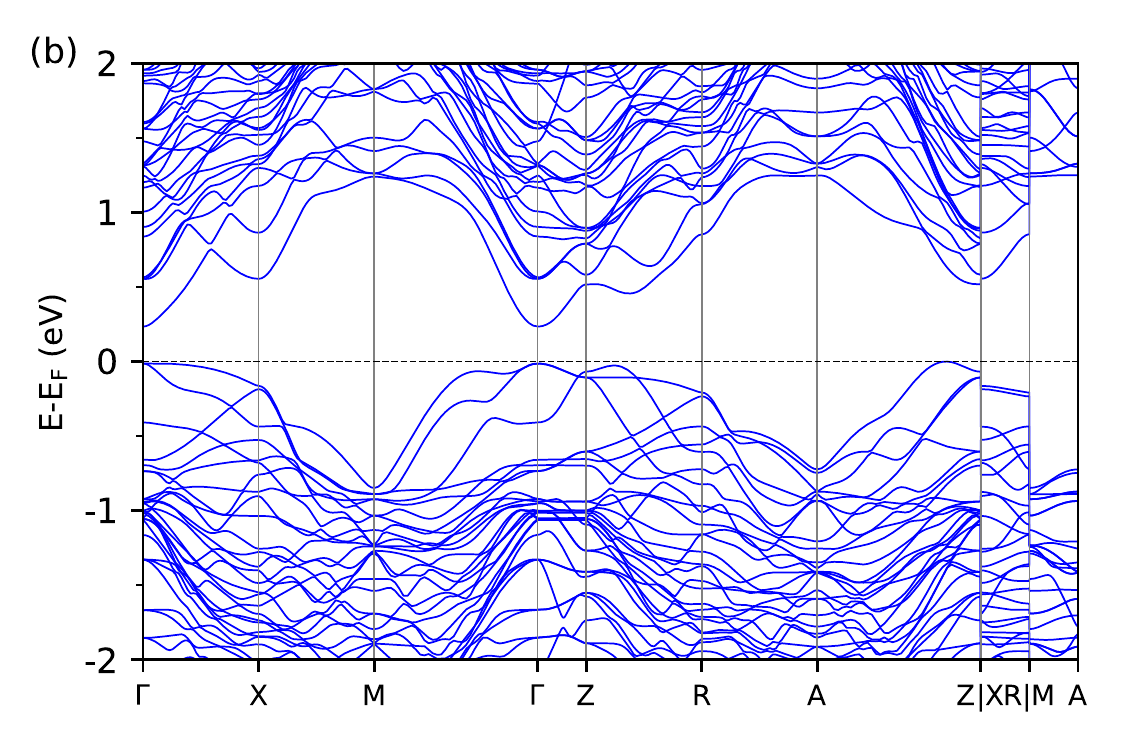}
\caption{The GGA-PBE calculated band structure of
(a) the 13-layer Ge Humble structure, and (b) the 13-layer
Ge$_{0.8}$Si$_{0.2}$ Humble structure. The dashed horizontal line
marks the Fermi energy, which is set at the valence bands maximum.}
\figlab{ge-gesi-humble-band}
\end{figure}

In Appendix~\ref{9-layer-humble-band}, we report the HSE03-calculated
band structures of the Ge and Ge$_{0.8}$Si$_{0.2}$
Humble structures. However, due to the computational
expense, these calculations were performed on a 9-layer Humble
structures instead of the 13-layer ones used here.
We again find that the HSE03 correction to the GGA-PBE band
structure is of the scissors type, so that the DOS
of the conduction bands calculated within the HSE03 approximation is
very close to those given by GGA-PBE after a rigid shift in energy.
This suggests that the DOS calculated using the
GGA-PBE can be used to simulate the experimental EEL spectra,
an expectation that is borne out by the good agreement between
theory and experiment that is presented below.

%=================================================
\section{EEL spectra of the Humble defect: experiments and simulations}
\seclab{eels}
%=================================================

In the past, core-loss EELS has mainly been used to
differentiate between different chemical bonding environments
of a given element~\cite{rez2008}. Here, we use Si L$_{2,3}$-edge
spectroscopy to distinguish between defected and bulk-like regions
of a Ge$_{0.8}$Si$_{0.2}$ sample, even though all the atoms are
four-fold coordinated.  We work with an instrumental energy
resolution of $100$~meV, noting that under these conditions
the EELS spectra are still limited mainly by the $2p$ core-hole
lifetime~\cite{klima1970,altarelli1972}.
After deconvoluting to obtain the L$_3$ edge spectra, the resolution
is sufficient for a direct comparison with theory.
The EEL spectra are measured in the bulk region and at the defect core. As discussed in Sec.~\ref{sec:method-eels}, the defect core spectra are measured separately for electron beam subscan windows with heights of ${3\,\mbox{\AA}}$ and $6\,\mbox{\AA}$ as indicated in \fref{window-bulk}(a).

\begin{figure}
\centering\includegraphics[width=\columnwidth]{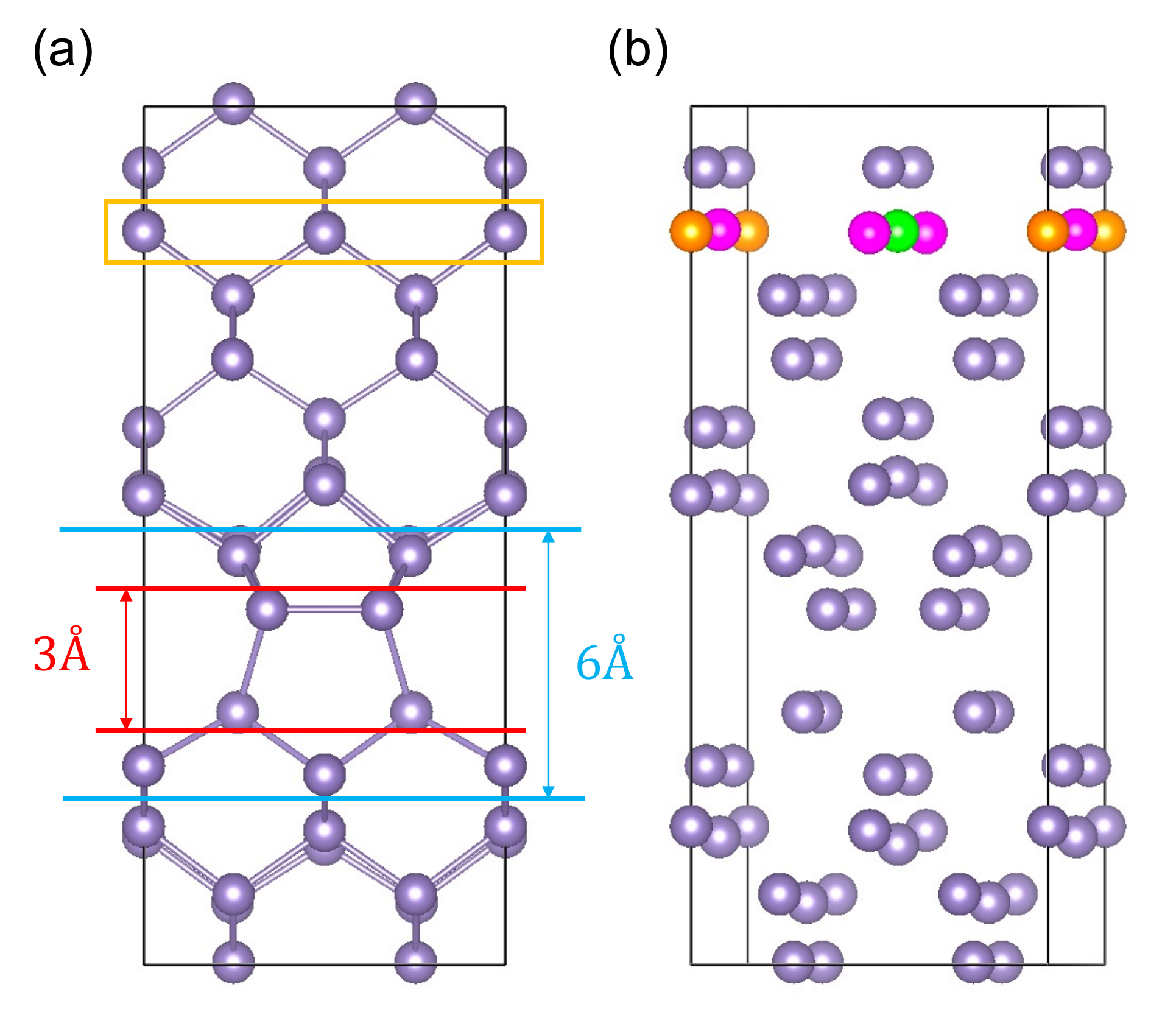}
\caption{(a) Side view of the unit cell of the 13-layer Humble
structure. The $3\,\mbox{\AA}$ window around the defect core is
shown as the red interval; the $6\,\mbox{\AA}$ window is shown as
the blue interval; the (most) bulk-like layer is the layer boxed
by the orange rectangle. (b) Different equivalent sites in the
bulk-like layer. The corner, edge, and center sites are in orange,
magenta, and green, respectively.}
\figlab{window-bulk}
\end{figure}

In the EELS measurement, the differential cross section
${d^2 \sigma}/{d\Omega d E}$ for electron scattering is given
by~\cite{weng1989}
\beq
\frac{d^2 \sigma}{d\Omega d E} = \frac{4\gamma^2}{a_0^2 q^2}\big{[}\vert m_{L+1}\vert^2\rho_{L+1}(E)~+~
\vert m_{L-1}\vert^2\rho_{L-1}(E)\big{]}\,,
\eqlab{cross-sec}
\eeq
where $\gamma$ is the Lorentz factor, $a_0$ is the Bohr radius, $q$
is the momentum transfer, $m_{L\pm 1} = \me{f_{L \pm 1}}{r}{i_L}$
is the electric-dipole transition matrix element slowly varying
with energy, $\ket{i_L}$ is the initial core-level state,
$\ket{f_{L \pm 1}}$ is the final conduction state, and $\rho_{L
\pm 1}(E)$ is the angular-momentum-resolved DOS.  Here we have
assumed the dipole selection rule ($\Delta L = \pm 1$) for the
transitions from the core-level to the conduction states.  For the
Si L$_3$ edges, since the $\ket{i_L}$ is the Si $2p$ core-level state,
the EELS is only sensitive to final states $\ket{f_{L \pm 1}}$
of predominant $s$ or $d$ character.

As mentioned in \sref{humble-ge-gesi}, the Humble structure in Ge
and Ge$_{0.8}$Si$_{0.2}$ have very similar electronic structures.
Considering the similarity
between Si and Ge atoms and the fact that an average of 80\% of the
neighbors of any given Si site are Ge atoms, we now abandon the
use of the VCA and work instead in the limit of low Si concentration.
That is, we carry out calculations using supercells in which only
a single Si atom has been substituted into the Ge Humble structure,
and vary the location of this impurity atom to take statistical
averages.

A further complication is that a core hole
is formed when a core electron is ejected,
giving rise to an interaction with the conduction bands.
In most cases, such a ``core-hole effect'' cannot be neglected,
especially in insulators~\cite{duscher2001,rez2008}.
Accordingly, we adopt the ``Z+1 approximation," in which the excitation
is simulated with an extra proton in the nucleus of the excited Si
atom~\cite{lee1977,buczko2000,duscher2001,rez2008,zhou2012}.
Using the structure as obtained from a relaxation without the
core hole,
we replace the Si atom with a P atom and then calculate
the $s$- and $d$-projected local DOS on
the P atom.
Typically, a large supercell is needed to avoid interactions
between periodic images of the Z+1 atoms.  We have tested
supercells having  52, 104, and 208 atoms, and find that
good convergence is achieved for
the 104-atom cell.  Conversely, if the core-hole effect is
negligible or fully screened, one can use the local $s$- and
$d$-projected DOS for an ordinary Si impurity; we refer to this as
the ``Z approximation"~\cite{weng1989,weng1990,rez2008}.

The spin-orbit coupling is included while implementing
the Z and Z+1 approximations. To save computational expense,
these supercell calculations are carried out using
GGA-PBE rather than HSE03. In
Appendix~\ref{9-layer-humble-band}, we show that the
correction coming from HSE03 is again mainly of the scissors type. That is,
the DOS of the conduction bands calculated using GGA-PBE and HSE03
are very similar after a rigid upward shift of the GGA-PBE calculated
conduction bands.  As will be shown below, we find that the
GGA-PBE simulations give good agreement with our experimental
EELS data.

All the reported EELS simulations are performed in 13-layer
Humble structures.
We include enough conduction bands so that all
states up to 8\,eV above the valence-band maximum are included.
For the calculation of the partial DOS, the considered Wigner-Seitz radii for
Si and P are 1.312 and 1.233\,\AA, respectively.
The DOS is broadened by a Gaussian function with a width
of 0.05\,eV.

%=================================================
\subsection{Bulk-like region}
\seclab{eels-bulk}
%=================================================

\begin{figure}
\centering\includegraphics[width=\columnwidth]{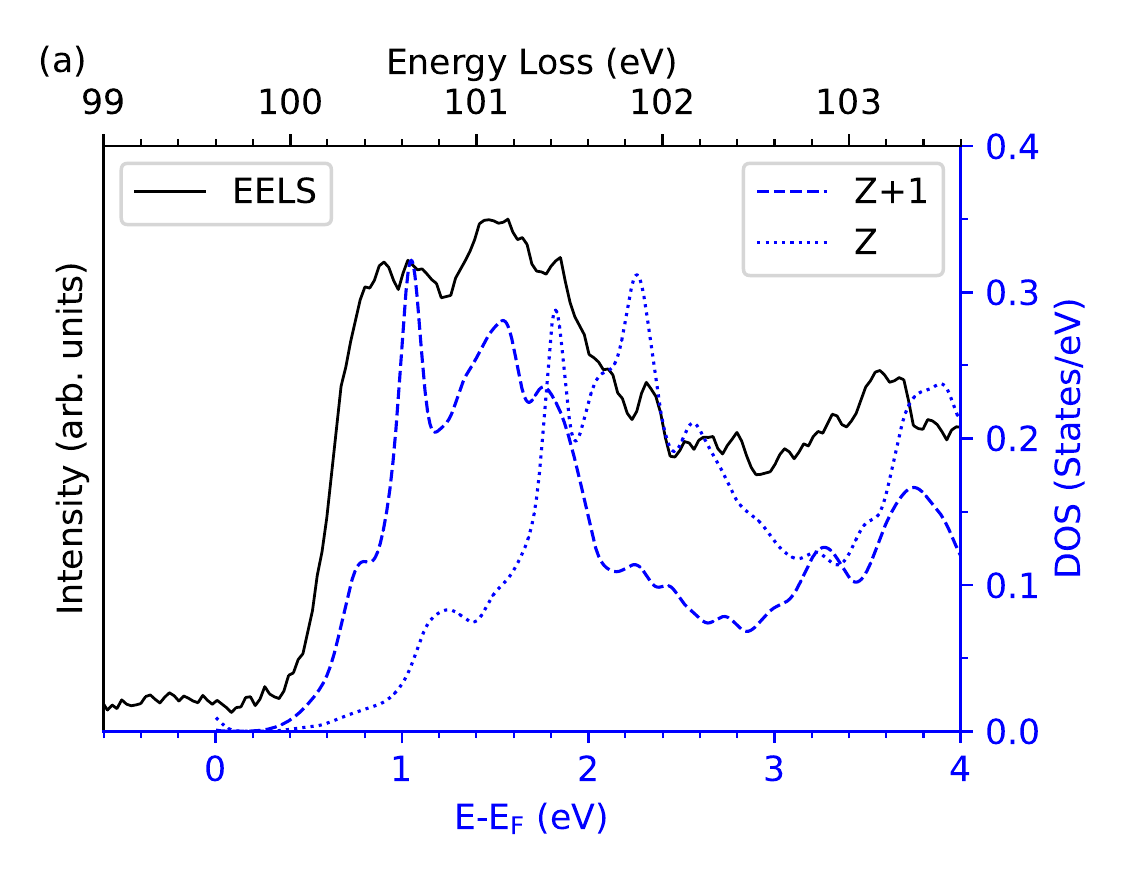}
\centering\includegraphics[width=\columnwidth]{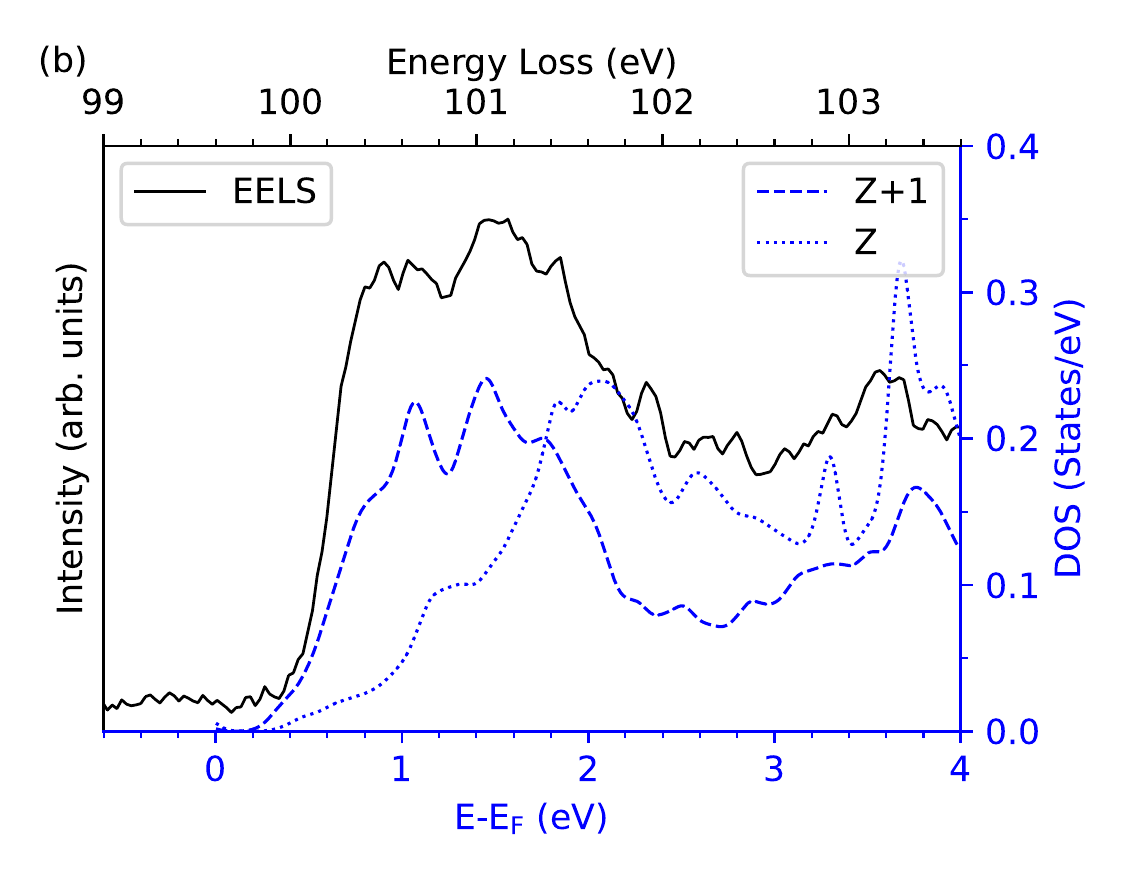}
\caption{Comparison of measured and computed Si L$_3$ edge
EEL spectra in the bulk region. (a) Simulated with the Si atom located at the
lowest energy site. (b) Same, but with the Si atom distributed
uniformly over all four sites. In each panel, the solid black line is the
experimental measurement; the blue dashed and dotted lines show the simulated
EEL spectra using the Z+1 and Z approximations, respectively.}
\figlab{eels-bulk}
\end{figure}

In this section, we present our results for the EELS measurements
and simulations in the bulk-like region of the Humble structure,
i.e., far from the Humble defect layer.
The EEL spectrum in the bulk-region is measured 1~nm away from
the defect core, and is essentially identical if measured 3~nm
away. For the simulations, the most bulk-like atoms in the
13-layer Humble structure are the four atoms
in the sixth atomic layer above or
below the defect core, as shown in \fref{window-bulk}.  We carry
out two sets of calculations, one with
the Si atom located at the most energetically favorable of the
four sites, and the other assuming a uniformly distributed over
all four sites.

For the first case, we determine the lowest-energy
site for the substitution of Si atom by computing the energy
cost of the substitution at each site.  By symmetry
there are three inequivalent sites, denoted as the
corner, edge, and center sites in \fref{window-bulk}(b),
We find the corner site to be most favorable, with the edge
and center sites are higher by 57 and 96\,meV respectively.
Then we apply both the Z and Z+1 approximation to the Si atom at this site to simulate
the EEL spectrum measured in the bulk region.  The experimental
data are compared with the theoretical results in
\fref{eels-bulk}(a).

For the second case, we obtain the simulated EEL spectrum $n_i(E)$
at each of the three unique sites separately, and average them as
\beq
n_{\rm ave} (E) = \sum_{i=1}^3 w_i n_i (E)\,
\eqlab{nall}
\eeq
with weights $w_i=0.25$ for corner and center sites and 0.5 for
the edge sites. The resulting Z and Z+1 spectra are shown in
\fref{eels-bulk}(b).

From \fref{eels-bulk}, it is evident that the Z+1 approximation
is in very good agreement with the EEL spectrum no matter which of these
two approaches we adopt.
On the other hand, the Z approximation fails to predict the peaks below
2\,eV, but it still provides some information about peaks
above 2\,eV.  This indicates that the core-hole effects cannot
be neglected while simulating the EELS measured in the
bulk region.

In the simulations described above, we only substitute one Ge atom
at a time by Si (or P) in the Humble
supercell.  In other words, no nearest-neighbor Si-Si bonds
are considered.
In Appendix~\ref{si-si}, we investigate
the effect of such Si-Si bonds by substituting two Ge atoms
simultaneously. The results show that Si-Si bonds do not
change the excitation spectra substantially, suggesting that
the single-impurity approximation is sufficient to model the
behavior of Si atoms
in the Ge$_{0.8}$Si$_{0.2}$ Humble structure.

%=================================================
\subsection{Defect region}
\seclab{eels-defect}
%=================================================

\begin{figure}
\centering\includegraphics[width=\columnwidth]{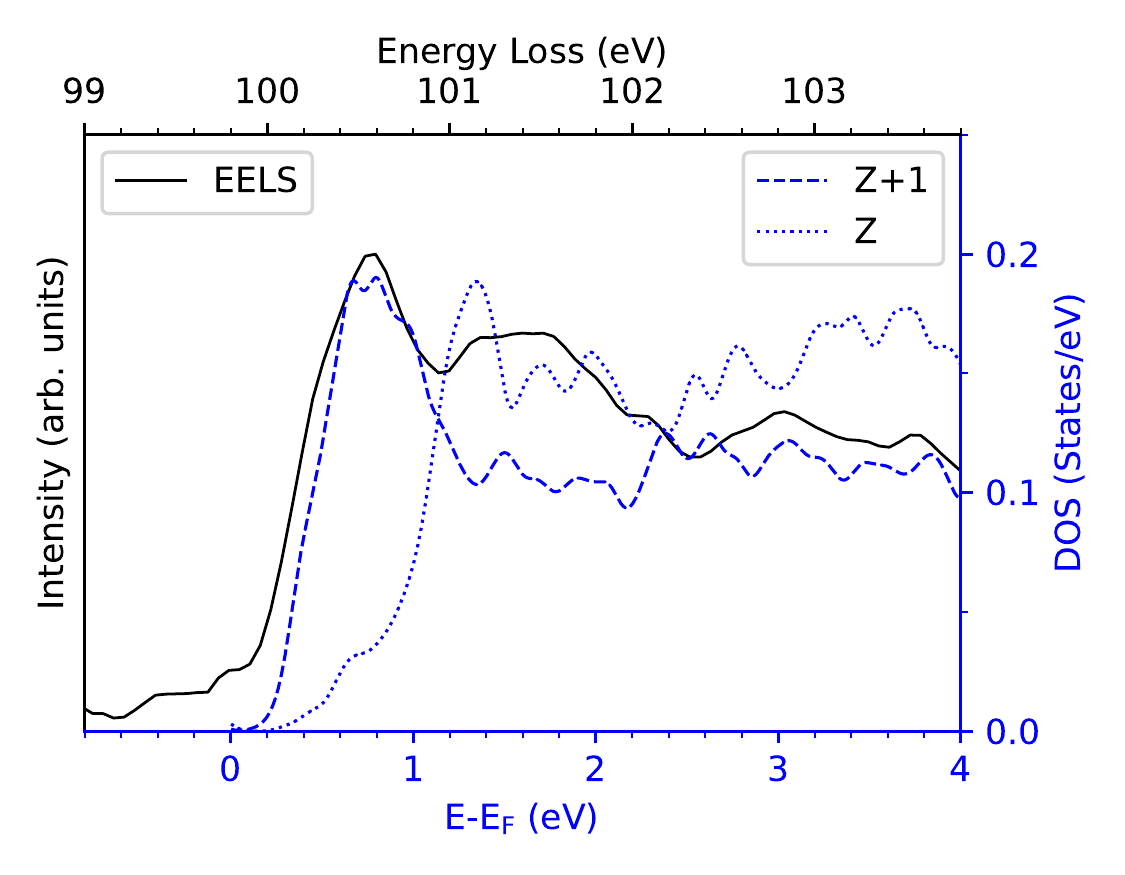}
\caption{Comparison of measured and computed Si L$_3$ edge
EEL spectra in the defect core (3\,\AA\ window).
The solid black line is the
experimental measurement; the blue dashed and dotted lines show the simulated
EEL spectra using the Z+1 and Z approximations, respectively.}
\figlab{eels-defect-3a}
\end{figure}

In the previous subsection, we used the Z and Z+1 approximations
to simulate the EEL spectrum and compared it with measurements in
the bulk-like region of the sample. Here we present similar comparisons,
but for the defect region.

In the defect core, EEL spectra are separately measured within
3\,$\mbox{\AA}$ and a 6\,$\mbox{\AA}$ integration windows.
Both windows are centered around the defect core, as
shown in \fref{window-bulk}(a).  The 3\,$\mbox{\AA}$ window
spans only the defect core, as shown by the red and blue
atoms in \fref{lang-humble-bz}(b), while the 6\,$\mbox{\AA}$
window 
includes two more layers adjacent to the core. The experimental
EEL spectra and the theoretical simulations, are shown in
\frefs{eels-defect-3a}{eels-defect-6a}.

By comparing \fref{eels-bulk} with
\frefs{eels-defect-3a}{eels-defect-6a}, it is clear that
the EEL spectra measured in the bulk and defect regions are
significantly different, although all Si atoms are four-fold
coordinated in both regions.  This suggests that some more subtle
difference in local bonding configuration must be responsible.
Similar differences are expected in the comparison of the spectra for the
different spatial windows defined in \fref{eels-defect-3a} and
\fref{eels-defect-6a}, since the
6\,$\mbox{\AA}$ window in \fref{eels-defect-6a} also covers some
atoms outside the defect core.

In the spectrum measured within
the  $3\,\mbox{\AA}$ window,
the intensity of the peak below $1$\,eV is relatively higher, when compared to the intensity of the adjacent peak lying in the $1$-$2$\,eV range of the same spectrum, than that of in the $6\,\mbox{\AA}$-window spectrum.  
This suggests that
the highest peak below 1\,eV might serve as a fingerprint of
the Humble defect.

Next, we compare our experimental measurements with the theoretical
simulations.  From \frefs{eels-defect-3a}{eels-defect-6a}, we
find that the Z+1 approximation works well in both cases,
although the peak intensities between 1\,eV to 2\,eV are slightly
underestimated. On the other hand, the Z approximation does not
work very well near the edge onset,
which indicates that the core-hole effect cannot be neglected in this system.

\begin{figure}
\centering\includegraphics[width=\columnwidth]{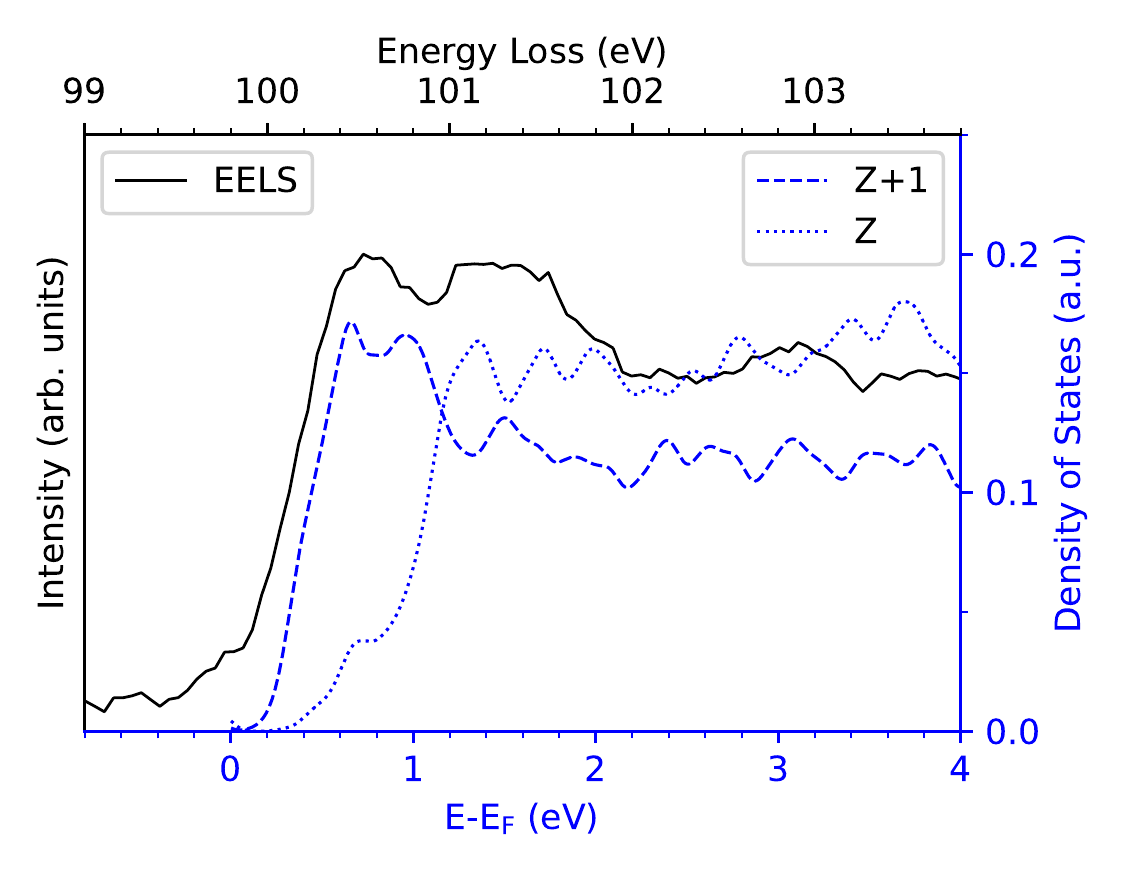}
\caption{Comparison of measured and computed Si L$_3$ edge
EEL spectra in the defect core (6\,\AA\ window).
The black line is the experimental measurement.
The solid black line is the
experimental measurement; the blue dashed and dotted lines show the simulated
EEL spectra using the Z+1 and Z approximations, respectively.}
\figlab{eels-defect-6a}
\end{figure}

%=================================================
\section{Discussion and conclusion}
\seclab{disscon}
%=================================================

In summary, we first calculated the band structures of
bulk Ge and Ge$_{0.8}$Si$_{0.2}$ at the GGA-PBE level. Although
both materials are insulating experimentally, LDA and GGA
calculations predict them to be semimetallic
due to the well-known band gap problem of DFT. This issue was
resolved by performing HSE03 hybrid-functional calculations,
which correctly predict band gaps that are in good agreement with
experiment.

Next, we used the GGA-PBE to calculate the electronic properties
of the Ge and Ge$_{0.8}$Si$_{0.2}$ Humble structures.  These are
both insulating in our calculations, although their defect-free bulk
structures are predicted to be semimetallic when calculated in the same
way.  This indicates that the Humble defect can locally enlarge the band gap of
Ge and Ge$_{0.8}$Si$_{0.2}$, a fact that may potentially be useful in band
engineering.
We also found that the DOS calculated using HSE03 is very nearly a rigid shift
of that calculated using GGA-PBE, justifying the use of the
GGA-PBE calculations for the simulation of the EEL spectra.

We have separately measured the Si L$_3$ edge EEL spectra in
Ge$_{0.8}$Si$_{0.2}$ in a bulk-like region and at the Humble defect.  To simulate
the EEL spectra, we used a single Si atom in the Ge Humble
structure to mimic the Si atom in the Ge$_{0.8}$Si$_{0.2}$
alloy, and we placed the Si atom at various sites to simulate
the EEL spectra measured from different regions. We think this is
a good approximation for three reasons. (i) The band-structure
calculations show that the Ge and Ge$_{0.8}$Si$_{0.2}$ Humble structures
have similar electronic properties. (ii)
The chemical properties of Si are similar to those of Ge, and the
concentration of Si is relatively low. (iii) The Si-Si bond
was not found to have a strong effect on the simulated EEL spectra.
We implemented both the Z and Z+1 approximations to simulate the EEL
spectra, corresponding to the absence and presence of the core
hole, respectively. The spectra simulated using the Z+1
approximation were in much better agreement with experiment, 
especially near the edge onset, indicating that
the core-hole effect is not negligible, as might be expected given
that the studied system is semiconducting so that no metallic screening
of the core-hole occurs.

However, the intensities of a few peaks were still not well predicted
by theory. There are several possible reasons. Firstly, the EEL
spectra are measured in the Ge$_{0.8}$Si$_{0.2}$ alloy.  Due to
the randomness of the alloy, the electron momentum $k$ is no
longer a good quantum number.  As a result, the dipole selection
rules in \eq{cross-sec} may be changed due to the disorder.
Secondly, the presence of some Si-Si bonds can affect the intensities of peaks.
This effect is discussed in Appendix.~\ref{si-si}, where
we find that a Si neighbor of the Si core hole has little effect on
the position of the peaks but does change their intensities. 
Thirdly, we have assumed that $\vert m_{L+1}\vert$ and
$\vert m_{L-1}\vert$ in \eq{cross-sec} are the same. However,
in Refs.~\cite{weng1989,weng1990}, it is reported that the
ratio of the intensities of the $p\rightarrow s$ and $p\rightarrow d$
transitions is about 2\,:\,1. We tried using
this ratio in the Z+1 calculation, but the ratio of 1\,:\,1 actually fits
better with the experimental measurements.  A proper determination of
this ratio would require use of an all-electron method~\cite{duscher2001}.
Finally, both the Z and Z+1 approximations are based on a
single-particle picture.  A more accurate treatment of the
electron-hole interaction could be carried out by solving the Bethe-Salpeter
equation~\cite{shirley1998,benedict1998,rohlfing1998}; this
could be a possible direction for further studies, but is beyond
the scope of the present work.

%=================================================
\section*{Acknowledgments}
%=================================================

First-principles DFT calculations were performed using the
Rutgers University Parallel Computing (RUPC) clusters. We
thank Viktor Oudovenko for his technical assistance. We also thank Emily Turner and Kevin S. Jones for preparing the sample. S.R. and
D.V. were supported by NSF Grant DMR-1954856. S.S. acknowledges the
support from the Office of Naval Research grant N00014-21-1-2107 and the U.S. Department of Energy (DOE), Office of Science, Basic Energy Sciences under award DE-SC0020353. 
H.Y. and E.G. acknowledge the  financial support from the US Department of Energy, under award number DE-EE0008083.

%=================================================
\appendix
\section{Band structures of Humble defects in Ge and
Ge$_{0.8}$Si$_{0.2}$: A hybrid functional study}
\label{9-layer-humble-band}
%=================================================

\begin{figure}
\centering
\includegraphics[width=\columnwidth]{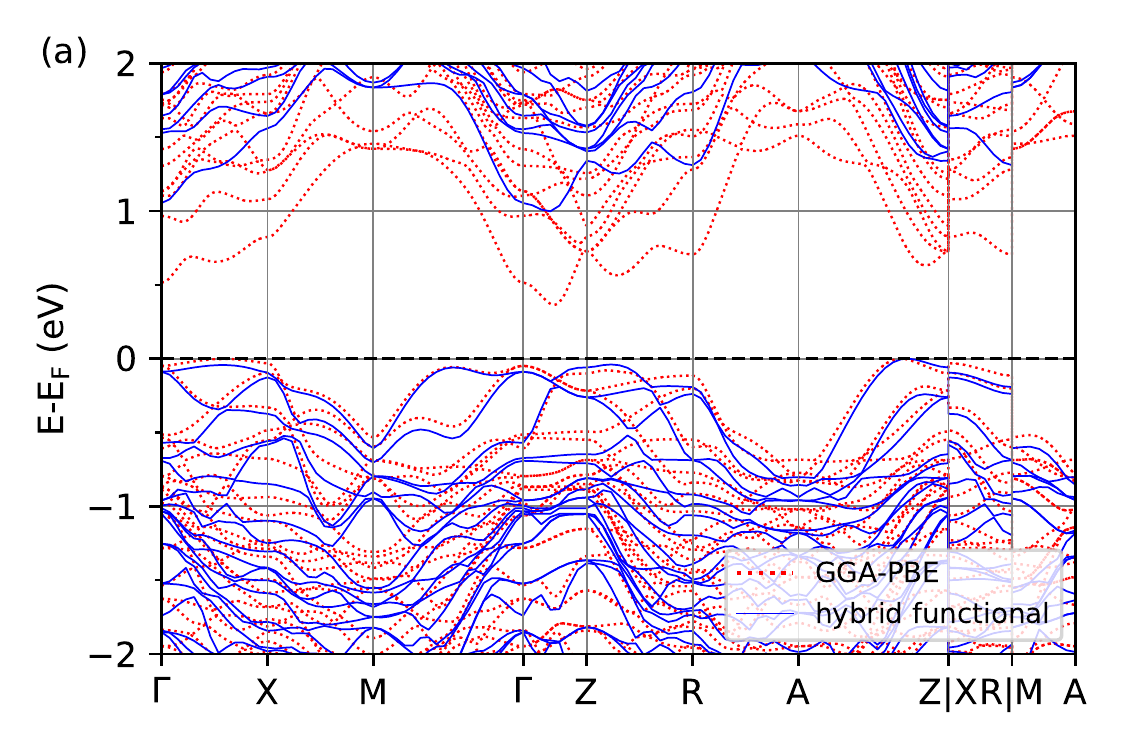}
\includegraphics[width=\columnwidth]{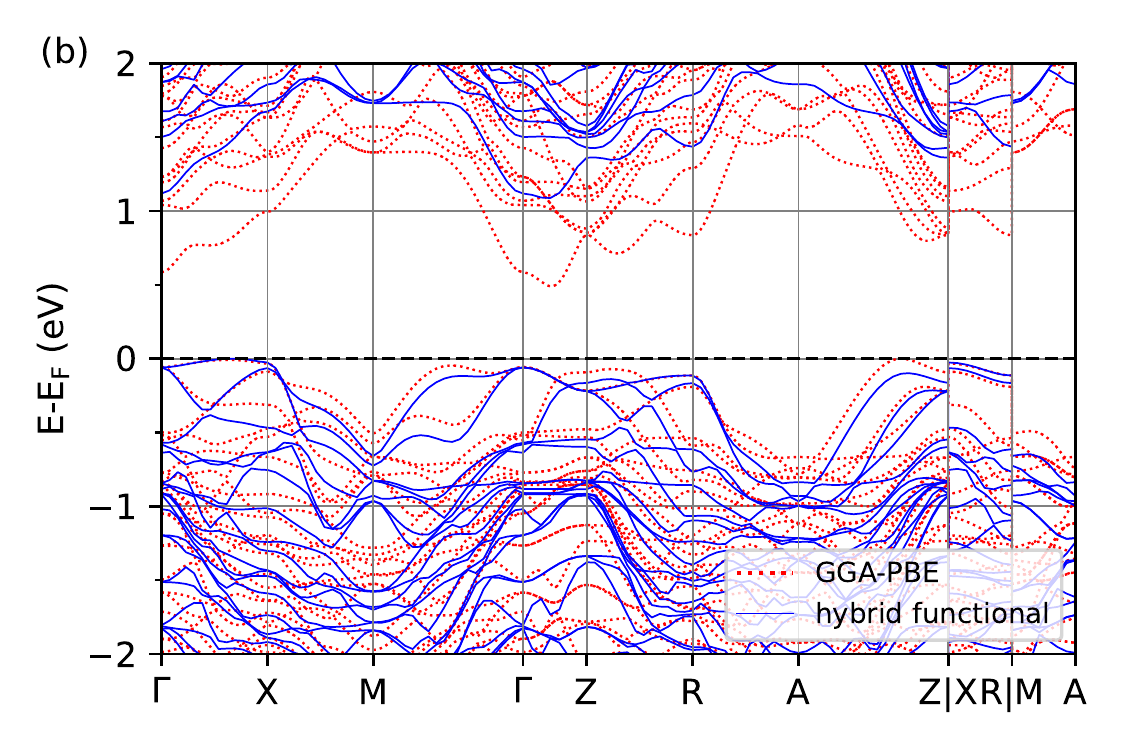}
\caption{Band structures of (a) Ge and (b) Ge$_{0.8}$Si$_{0.2}$
Humble structures. The dotted red and solid blue lines indicate the GGA-PBE
and HSE03 calculations respectively. The Fermi energy is at the top
of the valence bands.}
\figlab{band-dft-hf}
\end{figure}

\begin{table}
\caption{\label{tab:9-layer-ge-gesi} 
Band gaps of 9-layer Ge and Ge$_{0.8}$Si$_{0.2}$ Humble structures
computed using GGA-PBE and HSE03 at the specified lattice constants.}
\begin{ruledtabular}
\begin{tabular}{lccc}
& \multicolumn{2}{c}{Lattice Constants ($\mbox{\AA}$)} & Band Gap  \\
& $a,b$ & $c$ & (eV) \\
\colrule
Ge (GGA-PBE) & $8.18$ & $13.61$ & $0.36$  \\
Ge (HSE03) & $8.00$ & $13.27$ & $1.00$  \\
Ge$_{0.8}$Si$_{0.2}$ (GGA-PBE) & $8.16$ & $13.58$ & $0.49$ \\
Ge$_{0.8}$Si$_{0.2}$ (HSE03) & $7.93$ & $13.43$ & $1.09$ \\
\end{tabular}
\end{ruledtabular}
\end{table}

In this work, we have made use of two approximations. Firstly,
we use the DOS calculated from GGA-PBE instead of the more
accurate HSE03. Secondly, we use the Ge Humble structure
to approximate the Ge$_{0.8}$Si$_{0.2}$ Humble structure. In
this Appendix, we justify these two approximations by comparing
the band structures of 9-layer Ge and
Ge$_{0.8}$Si$_{0.2}$ Humble structures. We use both GGA-PBE and
HSE03 to calculate the bands for each structure.

\Fref{band-dft-hf} shows the band structures of the 9-layer Ge and
Ge$_{0.8}$Si$_{0.2}$ supercells in panels (a) and (b) respectively.
For each panel, the GGA-PBE bands, shown as dotted red lines, are calculated
for the DFT-relaxed structure as reported in Table.~\ref{tab:9-layer-ge-gesi}.
Although bulk
Ge and Ge$_{0.8}$Si$_{0.2}$ are semi-metallic if calculated
using the GGA-PBE, each Humble structure is insulating with an indirect band
gap.

The solid blue lines in \fref{band-dft-hf} show the HSE03 bands for the Ge and Ge$_{0.8}$Si$_{0.2}$ Humble structures respectively. To save computational cost, we use a $4\times4\times2$ MP grid of k-points for the self-consistent field calculations. As in \sref{bk-ge-gesi}, we use the experimental lattice constants when calculating HSE03 bands. However, because we do not have the experimental $c$ lattice constant for the Ge Humble structure, we determine it by fitting the stress $\sigma_{zz}$ to the value calculated for the bulk at the experimental lattice constant. The $c$ lattice constant for the Ge$_{0.8}$Si$_{0.2}$ Humble structure is measured experimentally.

By comparing panels (a) and (b) in \fref{band-dft-hf}, we again find that the Ge and Ge$_{0.8}$Si$_{0.2}$ Humble structures have similar band structures, and the momentum-space dispersion of GGA-PBE and HSE03 bands are very similar.

%=================================================
\section{The effect of Si-Si bonds in simulations of the EEL Spectra}
\label{si-si}
%=================================================

\begin{figure}
\centering\includegraphics[width=\columnwidth]{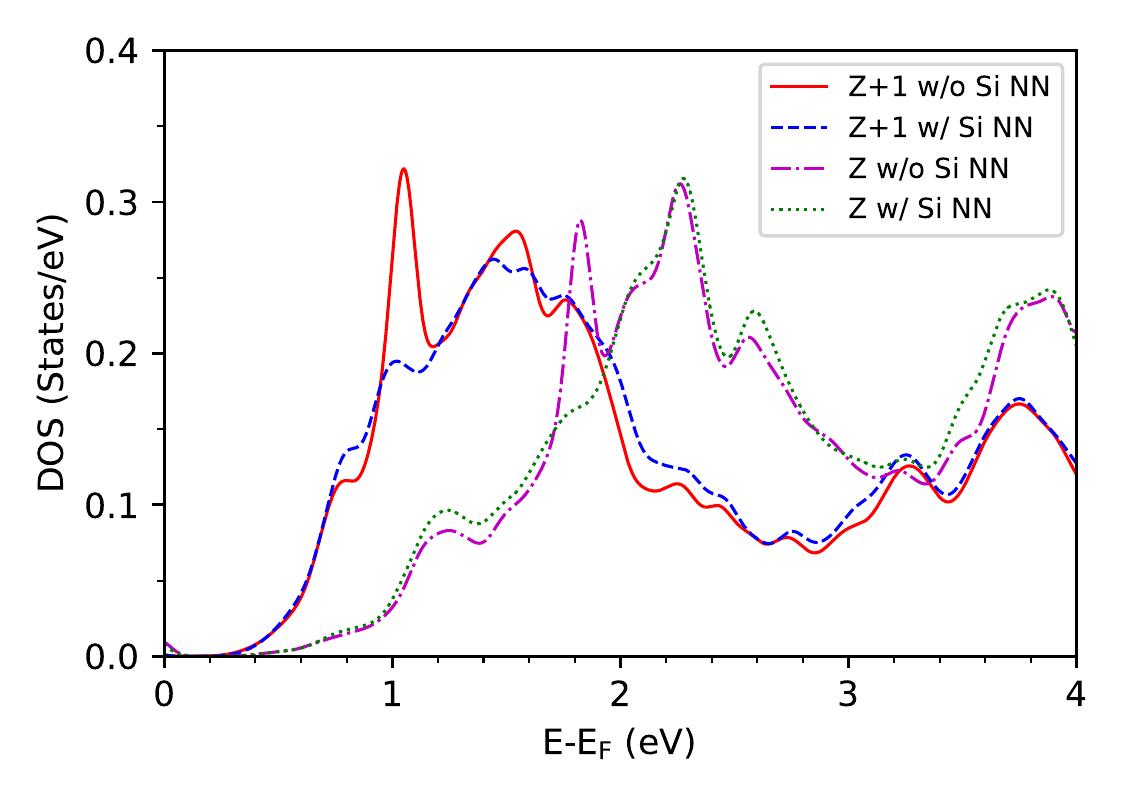}
\caption{Simulations of the Si L$_3$ edge EEL spectra in the bulk
region, with or without the Si nearest neighbor. The red solid
line and blue dashed line are simulated by the Z+1 approximation
without or with the Si nearest neighbor, respectively. The
magenta dash-dotted line and green dotted line are simulated
by the Z approximation without or with the Si nearest neighbor,
respectively.}
\figlab{w-wo-si-nn}
\end{figure}

In Appendix~\ref{9-layer-humble-band}, we have argued that we can
use the Ge Humble structure to approximate the Ge$_{0.8}$Si$_{0.2}$
one. We implement this approximation when simulating EEL spectra,
and we only replace one atom by Si in the Ge Humble structure. As
a result, no Si-Si bond is considered in the simulation. Due
to the low concentration, the Si atom tends to form bonds with
Ge atoms. Therefore, the effect of Si-Si bonds should be
minor. However, it is instructive to check how the Si-Si bond
affects the simulation of the EEL spectra.

Similar to what we have done in \sref{eels}, we use a
13-layer Ge Humble structure, but with two Ge atoms replaced by Si.
One of these is located at the corner site in the bulk-like
layer, as in \fref{eels-bulk}(a), and
another Si atom is its nearest neighbor. Then we implement the Z
and Z+1 approximations to the Si atom in the bulk-like layer. The
results are shown in \fref{w-wo-si-nn}. The cases without the
Si-Si bond are also included to show the difference.

As can be seen in \fref{w-wo-si-nn}, the Si-Si bond does not change
the results substantially for either approximation. Specifically, the
positions of peaks are unchanged, although the intensities of some
peaks do change. Considering the minor effect of the Si-Si bond,
we think the approximation of using a single Si atom in the Ge Humble structure
is justified when studying the Si core-level spectra in Ge$_{0.8}$Si$_{0.2}$.

\bibliography{pap}

\end{document}